\documentclass[11pt,twoside]{article}

\usepackage{asp2004}
\usepackage{epsf}
\usepackage{psfig}
\usepackage{lscape}

\begin{document}
\title{Multiplicity, activity and fast rotation in early-type stars}
\author{Ignacio Negueruela}  
\affil{DFISTS, EPSA, Universidad de Alicante, E-03080,
  Alicante, Spain}

\begin{abstract} 

There are obvious and direct ways in which the
presence of binary companions may affect activity in OB stars through
tidal interactions. In this review, however, I consider a more
fundamental role for multiplicity and explore claims that the
Be phenomenon may be intimately linked to binarity. I describe
the binary channel for the formation of Be stars 
and the ongoing discussion about the relative contribution of this
channel to the population of Be stars. I also present evidence
suggesting that some environments are more favourable for
the appearance of Be stars and explore whether this may be connected
to initial conditions, such as the chemical composition or the
distribution of rotational velocities on the ZAMS. 


\end{abstract}


\section{Introduction}   

As the presence of binary (or multiple) companions in a close orbit can
affect the evolution of a star, it is only natural to assume that it
can also have an effect on the occurrence of different kinds of
stellar activity. In some cases, the physical mechanisms connecting
the presence of a companion with activity are more or less
transparent. For example, the tidal force exerted by a companion can
excite different kinds of stellar oscillations, as discussed by
\citet{wil06}. Such induced oscillations are believed 
to occur in a few binaries containing OB stars \citep[e.g.,][]{quaint}.

Similarly, the tidal force exerted by a companion on the disk of a Be
star can result in photometric variability. For example, regularly
spaced, small brightenings of the B6\,IVe star HR 2968 have been
interpreted as due to the presence of a companion \citep{car99}. 
However, dedicated searches for periodic
photometric variability locked with the orbital period in Be/X-ray
binaries have rarely resulted in detections
\citep[cf.][]{roc93, clark99}, even though the
presence of a neutron star in a relatively close orbit suggests that
the tidal effect on the decretion disk would be rather strong in these
systems
\citep[but see][for examples of positive detections in SMC Be/X-ray
  binaries]{ce04}.  In some Be stars, (quasi-)periodic variations in
the shape of emission 
lines (V/R variability) occur with the orbital period of a companion
(see, e.g., Stefl et al.\ these proceedings). 
The presence of a companion is one obvious way of
providing a clock when a clear periodicity exists.



There are, however, deeper ways in which membership in a stellar
system may be connected to activity. Considering the focus of this
meeting on the use of active stars as 
laboratories to study the effects of fast rotation, I will centre on
Be stars and explore claims that binarity provides the mechanism to
spin up Be stars to high rotational speeds. Afterwards, I will
move to a higher level of multiplicity and discuss the likelihood that
membership in an open cluster affects the chances of becoming a Be
star. Finally, I will address the role that variations in metallicity
may have in the occurrence of the Be phenomenon.

\section{Be stars and critical rotation}
\label{sec:rot}

A relatively important fraction of moderately massive stars display,
at least at 
times, emission lines and a characteristic infrared excess which can
be explained in terms of a quasi-keplerian disk of material
surrounding an otherwise normal star. These are the Be stars
\citep[see][for an up-to-date review]{pr03}. Be stars
constitute the largest group of active OB stars. As a class, they are
characterised by rapid rotation and therefore represent an excellent
laboratory to study the effect of rapid rotation on the evolution of
OB stars.

Though emission lines are the most noticeable characteristic of Be
stars, we have to stress that not all emission-line stars with B
spectral type are Be stars.
For example, many interacting binaries display emission lines which
are due to flows of material between the two stars or {\it accretion}
disks around one of the components. Likewise, Herbig Be stars display
emission lines from a disk believed to be the remnant of the
accretion disk through which the star formed. Here we will restrict
ourselves to ``classical'' Be stars, those objects in which emission
is believed to arise from a disk of material expelled from the
star. Classical Be stars do occasionally lose their disks and are then
able to re-form them. We will take this capability to build the disk
from the inside as the defining characteristic of a Be star. I will
not discuss sgB[e] stars, as the role of rotation or binarity in their
activity is unclear, and a major monograph on these objects has just been
published \citep{km06}.

The actual incidence of the Be phenomenon is a matter of discussion
(see Section~\ref{sec:rule}). The usual practise is defining the Be/B
fraction as the number of Be stars in a population divided by the
total number of B-type stars (Be + non-Be). However, considering the
large difference in intrinsic brightness between early and late B-type
stars, it may happen that a magnitude-limited study of the Be star
population in some environment (for instance, an open cluster, or the
SMC) does not reach the faintest B-type stars. Because of this, it is
also customary to talk about the Be/B fraction down to a given
spectral type. This should be remembered when discussing the Be star
fraction in open clusters, as there is increasing evidence that Be
stars are much more numerous among stars evolved away from the Zero
Age Main Sequence (ZAMS)
than among stars still close to the ZAMS (see
Section~\ref{sec:evol}). For example, when considering a massive
well-populated cluster like NGC~663 \citep[cf.][]{pig01}, the Be/B
fraction is very high if we just count stars earlier than B5, but --
given the shape of the IMF -- may be rather lower if all the late
B-type stars are included.

\subsection{Rotation, how close to critical?}
\label{sec:crit}

The existence of a connection between fast rotation and the Be
phenomenon has been widely accepted for a long time
\citep[cf.][]{str31}. Be stars are observed to rotate at very high
speeds and even these high measured rotational
velocities represent an underestimation of the actual rotational speed
$\Omega$ because of the effects of
gravity darkening \citep{tow04,fre05}.

Fast rotation reduces the effective gravity on the outer
atmospheric layers, perhaps allowing material to be lifted out of the
star by effects that would otherwise be only minor disturbances.
Such lifting would be extremely easy if the outer layers of the star
were rotating at the critical speed $\Omega_{{\rm crit}}$, at which
their velocity would be comparable to the Keplerian velocity.
If Be stars rotate at $\Omega_{{\rm crit}}$ or {\it very} close to it,
then mass ejection can be explained by any mechanism that can supply a
small amount of energy or angular momentum to the material.

This idea has recently been explored by a number of authors. Among
them, \citet{kel01}, observing that (as discussed in
Section~\ref{sec:evol}) Be stars in open clusters are almost always
restricted to  
the upper region of their HR diagram, i.e., to relatively evolved
stars, and assuming that the development of the Be phenomenon is
intimately linked to fast rotation, interpret this fact as evidence
for an evolutionary effect. They note that the evolutionary
models of \citet{mem00} for stars in the range
$\sim 5M_{\sun} - 15M_{\sun}$ predict that the ratio
$\Omega/\Omega_{{\rm crit}}$ will increase as the stars evolve. As a
consequence, even though a few very fast rotators
may become Be stars while still close to the ZAMS, most relatively
fast-rotating stars will only develop Be characteristics 
during their latest stages in the MS, as they approach $\Omega/\Omega_{{\rm
    crit}}\approx1$.

If we could assume that all Be stars rotate at $\Omega/\Omega_{{\rm
    crit}}\approx1$, it would be very easy to explain the Be
    phenomenon. However, observational evidence against this
    hypothesis remains strong. After correcting for the effects of
    gravity darkening, 
\citet{fre05} conclude that, on average, Be stars rotate at $\Omega =
0.88 \Omega_{{\rm crit}}$. However, this average hides large
variations. From a statistical analysis of the equatorial rotation
rates of classical Be stars, \citet{cran05} concludes that early Be
stars rotate at significantly subcritical speeds. It can be argued that
$\Omega_{{\rm crit}}$ is perhaps not the physical magnitude relevant
to this problem, but still it seems clear that early B-type stars in
general and early Be stars in particular seem to rotate at
significantly lower $\Omega/\Omega_{{\rm
    crit}}$ rates than their late-type counterparts. As a
matter of fact, the rotational velocity distributions for early and
late B-type stars have markedly different shapes \citep{abt02}, with
a much higher fraction of slow rotators among early B-type stars.

\section{The binary channel for the formation of Be stars}
\label{sec:bin}

\subsection{Be stars in binaries}

\citet{har02} have proposed that
all Be stars are members of binary systems and that their disks are
built from material lost to the gravitational pull of the
companion. Such model is not accepted as a general explanation, but
the existence of Be/X-ray binaries (X-ray sources composed of a Be star
orbited by a 
neutron star) demonstrates that binarity may indeed play
a role (if rather different) in the origin of the Be phenomenon: some
Be stars have at a previous stage received mass (and angular momentum)
from a binary
companion. 

According to the standard model, most (if not all) Be/X-ray binaries
have formed via the 
same standard evolutionary channel. The progenitor is an 
intermediate-mass close binary with moderate mass ratio $q\ga 0.5$. 
The original primary starts transferring mass to its companion when it
swells after
the end of the hydrogen core burning phase (case B). Mass transfer results in a
helium star and a rejuvenated main sequence (MS) star. 
If the helium star is massive enough, it 
will undergo a supernova explosion and become a neutron star. If the
binary is not disrupted, it may emerge as a Be/X-ray binary.

This model has been developed in a number of references, among
which, \citet{hab87}, \citet{pol91}, \citet{por95} and
\citet{vbv97}. All these models assume that
the original primary must have a mass $\ga 12\, M_{\sun}$
(because otherwise it would not produce a neutron star) and that 
the original secondary receives an important amount of mass and
angular momentum from its companion. There is an implicit, but
fundamental, assumption hidden in this argument: the Be nature of 
the original secondary is due to accretion of 
high-angular-momentum material from the primary. However, there has
been no attempt at explaining how this is achieved. Of course, if
near-critical rotation is a {\it sufficient} condition for having a Be star,
then there is no need to provide a mechanism, because it has been
amply demonstrated that accretion of even a small amount of material (on
the order of 0.1$M_{\sun}$) is enough to spin up the star to critical
rotational velocity \citep[e.g.,][]{pac81}. 

Be stars in Be/X-ray binaries do not seem different from isolated Be
stars in any fundamental way. Ample evidence has conclusively shown
that their disks may disappear and then reform from the inside
\citep[e.g.,][]{tel98,neg01}. As they have likely accreted a large
amount of matter, their internal structure may have been slightly
affected \citep[cf.][]{bl95}, but their observed characteristics
differ from those of isolated Be stars only in ways that can be perfectly
explained by the effect of the orbiting neutron star on the
circumstellar disk. As detailed by  \citet{on01}, the presence of a
neutron star companion results in the truncation of the decretion disk
and the accumulation of material inside the truncation radius. This
leads to higher densities than in the disks of isolated Be stars, a
fact that seems to be confirmed by observations
\citep{rei97,zam01}. Okazaki \& Hayasaki (these proceedings) provide a
review of our understanding of the complex interaction between the Be
envelope and the neutron star. 

While Be/X-ray binaries are relatively easy to find, because of their
hard X-ray emission, they are not the only examples of Be stars formed
through mass transfer. As discussed below, other Be stars are believed
to correspond to different stages in this evolutionary path.
Moreover, if the predictions of 
population synthesis models are correct, many Be stars formed by mass
transfer in a binary may not be detectable as binaries, either because
they have eventually lost the remnant of the original primary, as may
happen in a supernova explosion, or because the mass ratio is large
and the effect of the low-mass companion on the spectrum of the Be
star will be small and very difficult to observe.

\subsection{The rule or the exception?}
\label{sec:rule}

Be/X-ray binaries are very obvious representatives of the fact that Be
stars can be produced via binary evolution. The theoretical
evolutionary path developed to explain their formation is reinforced
by observations showing interacting binaries that can be identified
with different stages along this path. Just to quote a few examples
from the literature, {\bf V380 Cyg} is a binary with a B1.5\,II-III
(14$M_{\sun}$, 17$R_{\sun}$) 
primary close to filling its Roche lobe and a B2\,V (8$M_{\sun}$,4$R_{\sun}$) 
secondary \citep{hb84}. This is an example of pre-contact system that
has not started mass transfer yet. {\bf RY Per} is a binary undergoing
mass transfer from an F7:\,II-III 
(1.6$M_{\sun}$, 8$R_{\sun}$) donor star onto a very fast rotating
B4:\,V companion (6$M_{\sun}$, 8$R_{\sun}$), which is already rather
more massive than the donor \citep{op97}. {\bf HD 45166} is composed
by a B7\,V star and a very unusual secondary of about 
the same mass, which \citet{so05} classify as quasi-WR. This could be an
example of an exposed He core. Finally, the B0.5\,Ve star
{\bf $\phi$ Persei} is one of the best known examples of a post-mass
transfer system, as it is
orbited by a low-mass sdO subdwarf, believed to be the
remnant of the original primary \citep{gie98}. A few other Be stars
are suspected of hosting subdwarf companions.

 Thus all the phases in the path to a Be/X-ray binary are well
documented. However, it is important to remember that, while in
Be/X-ray binaries the originally more massive star loses a high
fraction of its initial mass and is finally reduced to a compact
object, other evolutionary paths may involve mass transfer and result
in different outcomes. The mass donor does not need
to pass a substantial fraction of its mass to the mass gainer in order
to spin it to high $\Omega$. A system like AI~Cru, which contains 
a B2\,IVe (10$M_{\sun}$, 5$R_{\sun}$) primary and a B4:\,V secondary
(6$M_{\sun}$, 4$R_{\sun}$) is believed to have experienced a previous
phase of mass transfer in which only a fraction of the envelope of the
originally more massive star was transfered \citep{bel87}.

Summarising, observations clearly indicate that the binary
channel contributes to the formation of Be stars and we have reasons
to believe that the intrinsic properties of these Be stars are
indistinguishable from those of any other Be stars. A direct
application of  
Occam's razor would suggest that, once we have found a channel to
produce Be stars, we can safely assume that it is the only channel
needed. Again, this assumption fundamentally depends  on the issue of
critical rotation. If 
having a star rotating at $\Omega_{{\rm crit}}$ is a {\it sufficient}
condition for having a Be star, it does not really matter how this
critical rotation is achieved. If it is not, then we have not really
provided a mechanism for making Be stars, but simply assumed that
they form after mass transfer.

Additional indirect evidence supporting an important contribution of
the binary channel to the formation of Be stars has been presented by
\citet{gies01}. However, this is not a generally accepted idea,
because of two reasons.
On the one hand, there is little observational evidence to support the
idea that all (or even most) Be stars have binary companions.
On the other hand, modern
 population synthesis models predict that the fraction of stars that
 have been through a mass transfer phase is small (or very small)
 compared to the number of Be stars \citep[e.g.,][]{vbv97}. According
 to these models, $\sim 4\%$ of B-type stars should experience mass
 transfer in a binary and so an even smaller fraction are expected 
 to be Be stars formed through mass exchange. This is very far from
 the $\sim 10-20\%$ (depending on spectral type) of Be stars amongst
 B-type stars in the Bright Star Catalog \citep[cf.][]{zb97}. 

Against this, \citet{msg05}, based on a comprehensive photometric
study of Be stars in a large sample of open clusters, argue that the
true Be/B fraction in the Galaxy is $\la 5\%$ and speculate that the
high fraction among bright B-type stars may be due to a selection
effect related to the age of Gould's Belt. Moreover, they claim
that the position of Be  
stars in the HR diagrams of open clusters indicates that a
substantial fraction of them are created through binary evolution. 
 The discrepancy between the Be fraction in Galactic open clusters
 found by \citet{msg05} and the statistics of bright field Be stars
 may perhaps simply stem from
 the fact that \citet{msg05} only consider Be stars {\it at a given
   time}, while \citet{zb97} use the historical record. However,
 \citet{mar05} find a Be/B fraction $\sim 15-20\%$ for early B types
 in the LMC field,  
 forcing us to conclude that the true Be/B fraction is not 
 well known at this point and the ability of population
 synthesis models to reproduce it is still an open subject. 

In addition, all population synthesis models predict that a
significant fraction of the Be stars created through the binary
channel must contain a white dwarf (WD) as the remnant of the originally
more massive star \citep[e.g.,][]{vbv97,rag01}. In spite of dedicated
searches, so far no 
convincing candidate has been found for a Be + WD binary
\citep{mot06}. A certain number of Be stars are known to 
have low mass companions whose nature is uncertain due to the lack of
spectral signatures. Some of them might be WDs, but until definite
examples of Be + WD binaries are identified, population synthesis
models remain suspect.

\section{Rapid rotation in open clusters}
\label{sec:clus}

A number of works have provided evidence in the sense  
that OB stars in clusters rotate on average faster than those in the
field. Different explanations for this effect have been presented, some
of them invoking physical causes, with others trying to understand it
as a selection effect.

\citet{gh04} argue that
OB stars brake during their evolution, resulting in lower $\Omega$ for
evolved stars. If the field population is on average older than the clusters
surveyed, it may contain more slow rotators. Conversely, \citet{kel04}
argues that stars in clusters rotate faster because the bright members
generally observed are stars close to the turnoff
	undergoing spin-up at the end of their MS lifetime.
Other interpretations do not resort to evolutionary
hypotheses, but attribute this effect to the initial
conditions. \citet{gut84} speculated that perhaps the bulk of the
field population originated in low density regions in the outskirts of
OB associations, where the formation of slow rotators may be
favoured. 

A similar line of argument is followed by \citet{str05}. Recent work
has convincingly shown that very high accretion rates 
are necessary for the formation of massive stars
\citep{bm01,ys02}. These high accretion rates are easier to attain near
the cores of massive clusters, induced by high gas pressure
\citep{mkt03}. As a consequence, stars formed in massive clusters are
more likely to be born with high $\Omega$. In order to test this
hypothesis, \citet{str05} have measured the
rotational velocities for a large sample of stars in the massive twin
clusters h \& $\chi$ Persei and compared them to the rotational
velocities of a sample of field
stars selected on the basis that they
occupy the same positions in the $\beta$/$c_{0}$ diagram (and
consequently are in the same evolutionary stage) as the cluster stars.

The results of \citet{str05} have challenged
many long-held assumptions. They find that the distribution of rotational
velocities for mid and late 
B-types among members of h \& $\chi$ Persei is {\it very} different
from the distribution among field stars of the same age, mainly
because of an almost complete absence of slow rotators among cluster
members. The difference between the $\Omega$ distribution of cluster
and field stars decreases for earlier spectral types to the point that
it is not statistically significant for stars earlier than B2.

According to \citet{str05}, these results imply that 
fast rotation in clusters is primordial and not due to an
evolutionary effect, with mid- and late-B stars in clusters being born
rotating faster than field stars of the same mass. 
The observations are interpreted within
the context of the magneto-centrifugal model for pre-MS stars
\citep{shu94}. The rotational velocity of the protostar will be fixed
by the interaction between its accretion disk and its magnetic field. For a
given magnetic field, higher accretion rates will result in higher
rotational velocities, favouring fast rotators in massive
clusters. Moreover, mid-B stars have a large radius at 
the birth line because of the onset of deuterium shell burning and
spin up considerably during their descent to the ZAMS . 

The implications for early B stars  are not so
	clear. \citet{str05} take their observational result at face
	value and conclude that early B stars in clusters do not
	rotate faster than field stars of the same mass. However, it
	must be noted that all the stars in h 
	\& $\chi$ Persei with spectral types earlier than B2 are rather
	evolved and that a significant fraction of
	them appear to be blue stragglers \citep{mar06}. As models for
	angular momentum transport in fast rotators in the
	corresponding mass range are still being
	developed (cf. the non-trivial differences between
	models in \citealt{mem00} and \citealt{mem03}), this remains, in my
	view, an open issue.

\section{The evolutionary effect in Be stars}
\label{sec:evol}

The results of \citet{str05} are conclusive with respect to the
$\Omega$ distribution in h \& $\chi$ Persei, but seem difficult to
reconcile with the well established fact that Be stars in open
clusters are much more frequent among stars evolved away from the ZAMS
than among stars still close to the ZAMS. This effect was first
remarked by \citet{ft00}, who, based on data collected from the
literature for a 
small sample of clusters, concluded that Be
stars were most frequent in clusters with the MS turnoff (MSTO) in the
B1\,--\,B2 range (and, correspondingly, ages in the 12\,--\,24 Myr
range). 



\citet{kel99,kel00} carried out a survey of very populous LMC and SMC
clusters in search of Be stars with narrow-band H$\alpha$
filters. They found large numbers of Be
stars, with the highest proportion of Be stars occurring at the 
earliest spectral types. They also noted a tendency for a higher Be
star fraction near the MSTO in most clusters \citep[see
  also][]{joh02}, though they did not find a clear correlation between
cluster age and Be fraction. Later,
\citet{kel01} also found that most of the Be stars in h \& $\chi$
Persei were located close to MSTO. The concentration of Be stars close
to the MSTO is confirmed in the much larger sample of \citet{msg05}.

Based on this observational result, \citet{kel01} developed the 
evolutionary model for the Be phenomenon presented in
Section~\ref{sec:crit}, assuming a direct connection
between critical rotation and the onset of the Be phenomenon. As
discussed at length in \citet{neg04}, observational 
evidence shows that stars with at least $25M_{\sun}$ can develop the
Be phenomenon, but this does not invalidate the basic premise of
\citet{kel01}, as the newer models by \citet{mem03} show that
$\Omega/\Omega_{{\rm crit}}$ also increases during the lifetimes of
stars with masses in the $\sim 15M_{\sun} - 25M_{\sun}$ range.

\subsection{Facts and biases}

Unlike previous studies, \citet{msg05} have observed and analysed a
large and statistically significant sample of open clusters
using a homogeneous method. It is therefore very significant that
their conclusions are rather different from those of other
works. They find a much lower Be/B fraction than generally assumed and
do not find strong dependences of the Be fraction on several
parameters which were generally thought to be relevant. 

All these works are based on photometric searches for emission-line
stars. Photometric techniques are very efficient at detecting strong
Balmer-line emitters, but do not discriminate weak emitters very
well \citep[a detailed discussion is presented in][]{msg1}.
Because of this, they are prone to suffering from several important
selection effects. First, it is unlikely that observations may be
equally sensitive for early and late B-type stars (late 
B-type stars in a given cluster may not be surveyed at all if there is
  a magnitude limit). Moreover, early Be stars tend to be strong
  emitters, while late Be stars tend to be weak emitters. Both effects
  conspire to make detection of early Be stars rather easier than the
  detection of late Be stars.

Perhaps even more importantly, 
most of the evidence for the Be phenomenon as an evolutionary effect
is provided by observations of a few ``Be-rich'' clusters. Be stars
have been searched for in the most populous LMC and SMC clusters
\citep{kel99} and in the traditionally-called ``Be-rich'' clusters in
the Galaxy, namely, NGC 663 \citep{pig01}, h \& $\chi$
Persei \citep{kel01,bk02}, NGC 3766 \citep{msg1} and
NGC~7419 \citep{pig00}. This pre-selection of {\it interesting}
targets again introduces a discernible bias in the results expectable.

Most of these clusters were surveyed for Be stars precisely
because they were known to be rich in Be stars. They are not necessarily
  representative of the general population, and indeed the results of
\citet{msg05} suggest that they are not. Moreover, all the ``Be-rich''
cluster in the Galaxy and most of populous clusters observed in the
  Magellanic Clouds have ages such that
  the MSTO is located close to spectral type B2, perhaps explaining
  the conclusions of \citet{ft00}.


Clear demonstration of the danger of preconceived ideas comes from
the double cluster h \& $\chi$ Persei, generally
listed as having a population extremely rich in Be stars
\citep[see][and references therein]{ft00}. A complete
spectroscopic survey of h \& $\chi$ Persei finds that the fraction of
Be stars in these clusters  is $\sim13$\% among early B-type stars and
negligible for stars later than B4 \citep{bk02}. This is not a very high
fraction. It is indeed lower than many estimates of the Be fraction
for the field  population. The double cluster is
very massive \citep{sle02} and so the
high number of Be stars is simply a consequence of the high number of
(B-type) stars, not of a high Be/B fraction.

Other open clusters really have a high Be fraction among
stars around the MSTO. Classical examples are 
NGC~663 ($\sim25\:$Myr), showing $\sim33\%$ for stars earlier than B5 at a
given time, NGC~3766 ($\sim25\:$Myr), where a very high fraction of
early B-type stars have been seen at some point in a Be phase, and
NGC~7419, ($\sim15\:$Myr), with $>30\%$ among early B-type stars at a
given time (Negueruela et al., these proceedings). However, as
discussed above, these clusters may not be representative of
  the average young open cluster in the solar neighbourhood. For a
  start, these three clusters are relatively massive. This 
may not only have an implication on the initial stellar rotational
velocities (see Section~\ref{sec:clus}), but also means that the number
of B stars is high enough to make the Be/B fraction statistically
significant. 

In order to see the significance of a high number of members, let us
consider one of the best (if not the best) studied open clusters in
the sky, The Pleiades. This cluster has an estimated age of 130~Myr,
with the stars around the turnoff having spectral type B6\,IV.
Out of 15 B-type stars in this cluster, 4 are Be stars
\citep{al78}. At $\sim130\:$Myr, the three B9.5\,V
stars in the cluster cannot be considered to be very
evolved and we can calculate the Be/B fraction for stars close
to the MSTO using only spectral type B9 and earlier. This results in 4
Be stars
out of 12 B stars or 33\%, a percentage as high as in the ``Be-rich''
clusters. However, the number of stars is too low to accept this result as
statistically meaningful. The same happens in many other clusters in
the 100\,--\,200~Myr range, which have few B stars left,  meaning that
we are likely to accept as 
statistically significant only calculations for massive clusters,
which may not be representative of the general population.

In summary, though it seems clear that there is an evolutionary effect
such that the Be fraction appears much higher among the (moderately)
evolved stars in a population than among the unevolved members, the
nature and reasons for this effect cannot be ascertained. \citet{kel01}
have argued that it is due to the evolution of rotational velocity as
the stars age, but \citet{msg05} find that the location of Be stars
in the HR diagrams of open clusters does not favour this interpretation
and \citet{str05} have shown that the $\Omega$ distribution among
stars in h \& $\chi$ Persei is exactly the opposite from what this
hypothesis would require. Some other mechanism must be required to
complement (or completely replace) $\Omega$ evolution. \citet{msg05}
suggest that this may be mass transfer in a close binary, but again
this model is not without complications.

Finally, in spite of the evidence accumulated for the existence of
this evolutionary effect, there appears to be a non-negligible
fraction of Be star very close to the ZAMS (cf. 
Wisniewski et al., these proceedings, and see also
\citealt{zor05}). Wisniewski et al.\ find substantial numbers of Be
stars in clusters younger than 10~Myr old, while some good examples of Be
stars with early spectral types in open clusters in the
3\,--$\,10\:$Myr range exist. Such objects  are very likely
incompatible with mass transfer in a binary system. \citet{zor05} try
to reconstruct the evolutionary status of a sample of Be stars and
come to the conclusion that early B-type stars (with $M\ga12M_{\sun}$)
may display the phenomenon at any stage during their MS life, but
less massive stars only become Be stars after a significant fraction
($\sim0.5$) of their lifetime. Once again, if the
Be phenomenon may simply be equated with having $\Omega/\Omega_{{\rm
    crit}}\approx1$, many different channels may contribute to the
population. If this is not the relevant condition, the evolutionary
effect in Be stars is still very far from being understood.

\subsection{Metallicity and the Be phenomenon}
\label{sec:met}

From a study of data in the literature, \citet{mae99} found that the
fraction of Be stars increases with 
decreasing metallicity ($Z$). However, their work used the populous SMC
clusters as representative of the low-$Z$ environment, thus
introducing a likely bias. From a more homogeneous study,   
Wisniewski et al.\ (these proceedings) find a dependence of the Be
fraction with $Z$, though not as strong as suggested by
\citet{mae99}. \citet{msg05} find suggestive, but not compelling,
evidence for a higher Be fraction among clusters outside the Solar
Circle than among clusters inside it, again suggesting some sort of
weak dependence of the Be fraction on $Z$.

The reasons for this dependence are not obvious. Among massive stars,
lower $Z$ will result in weaker stellar winds. According to
evolutionary models \citep[e.g.,][]{mem00}, the stellar
wind is very efficient at removing angular momentum from a star of
solar metallicity, but not so much at SMC metallicities. As a result,
solar metallicity stars will spin down during their lifetime much more
significantly than lower $Z$ stars. This will translate into a
higher fraction of fast rotators. Indeed \citet{kel04} finds that stars
in a sample of LMC early-B objects rotate faster (at almost 2-$\sigma$
significance) than stars in a sample of early-B objects in Galactic
open clusters. However, as the higher Be fraction at lower $Z$
is not confined to the earliest spectral types, but seems to extend to
lower mass B stars, with luminosities incapable of sustaining a
radiative wind, the absence of wind braking cannot be the main cause
of the increased Be fraction at low $Z$. 

Perhaps the initial conditions for star formation result in higher
average $\Omega$ at lower $Z$, though the reasons for this remain
unexplored. In any case, the dependence of the Be fraction on $Z$
is not yet ascertained. \citet{mar05} find a comparable Be fraction among
the LMC field population and in the Solar neighbourhood. An even more
striking cautionary example is provided by two relatively massive
clusters in the Cas OB8 
association. This group consists of at least five open clusters known
to have similar ages and distance estimates, situated in a relatively
small area of the sky. As such, they are expected to have been formed
in a single episode of star formation, probably because of the
fragmentation of a giant molecular cloud.

The most massive cluster in Cas~OB8 is NGC~663 \citep{mar06}. A
spectroscopic survey of 150 likely members was 
conducted by \citet{neg05} in October 2002. At that time, about one
third of all stars earlier than B5\,V were in the Be phase, with
little variation between evolved (giants and subgiants) and MS stars. 
The second most massive cluster in Cas~OB8 is NGC~654. \citet{sh99}
carried out a spectroscopic survey observing more than 40 B-type
members down to spectral type B4\,V. Near the cluster core, where
all the objects down to a given magnitude were observed, they found  
2 Be stars out of 34 early-type B stars. The fact that two
clusters likely to have formed at the same time from the
same material have extremely different Be/B fractions
suggests that chemical composition is not the main factor affecting
the eventual appearance of a large population of Be stars.

\section{Conclusions}

Though the study of large populations of early type stars in open
clusters seems to be the only way to gain an understanding of the
causes and incidence of the Be phenomenon, data so far has provided a very
blurred and, in many aspects, contradictory picture

We know for certain that some fraction of Be stars must have undergone
mass transfer in a binary system and we have reasons to suspect that
their Be nature may be in some way due to this process. However,
different authors have provided completely divergent estimates of the
contribution of this binary channel to the formation of Be stars. In
my opinion, the lack of evidence for any companion in many Be stars,
together with the statistics of the Be fraction and star counts in
clusters, make it unlikely that most Be stars form through this
channel. On the other hand, I believe that present
evidence suggests that a non-negligible fraction of Be stars form in
this way.

Several works have shown that B-type stars in (at least massive) open
clusters rotate 
faster than those in the field. This fast rotation is likely to be
primordial and related to the initial conditions of star formation, as
seems to be confirmed by the results of \citet{str05}. It is, however,
interesting to note that in the case of h \& $\chi$ Persei, this
higher average $\Omega$ has not {\it yet} resulted in a high Be
fraction. 

There is abundant evidence demonstrating that Be
stars in open clusters are more numerous around the MSTO,
suggesting some sort of evolutionary effect, but claims of a preferred
age with a higher fraction of Be stars are likely based on biased
samples. There is also some evidence favouring an inverse dependence
of the Be fraction on metallicity. However, at this point we do not
really understand which mechanism(s) may cause this dependence.  
The fact that the fraction of Be stars varies enormously from cluster
to cluster, even at a given $Z$, strongly suggests that the
conditions leading to the development of the Be phenomenon must be
primordial and have to be found at the cluster level.

\acknowledgements 
This research is partially supported by the Spanish
  Ministerio de Educaci\'on y Ciencia under grant AYA2002-00814 and
  the Generalitat Valenciana under 
  grant GV04B/729. The author is a researcher of the
programme {\em Ram\'on y Cajal}, funded by the MEC and the University
of Alicante. I thank J.~S.~Clark and J.~M.~Torrej\'on for comments on
this paper, which I would like to dedicate to the memory of
John~M.~Porter. This 
research has made use of the NASA's ADS Abstract 
Service, the Simbad database, operated at CDS,
Strasbourg (France) and the WEBDA open cluster database.


\end{document}